\documentclass[reprint,twocolumn,superscriptaddress,showpacs,preprintnumbers,nofootinbib,aps,prd,floatfix]{revtex4-1}

\usepackage{graphicx}
\usepackage{amsmath,amssymb,amsfonts}
\usepackage{stackrel}
\usepackage{enumitem}
\usepackage[english]{babel}
\usepackage{verbatim}
\usepackage{comment}
\usepackage[normalem]{ulem}
\usepackage{xcolor}
\usepackage{dcolumn}
\usepackage{physics}
\usepackage{multirow}
\usepackage{makecell}
\usepackage{bm}
\usepackage{slashed}
\usepackage{float}
\usepackage{eso-pic}
\usepackage{placeins}

\definecolor{darkred}{rgb}{.7,.1,.1}
\definecolor{dark-green}{rgb}{0.1,0.7,0.3}
\definecolor{darkyellow}{rgb}{0.75,0.55,0.0}

\newcommand{\tf}{\texorpdfstring}
\newcommand{\gev}{~\text{GeV}}

\def\nn{\nonumber}

\graphicspath{{./}{figures/}}

\usepackage{hyperref}
\hypersetup{
    pdfnewwindow=true,
    colorlinks=true,      
    linkcolor=darkred,
    citecolor=blue,        
    filecolor=blue,
    urlcolor=blue,
    pdfborder={0 0 0},    
    pdfencoding=unicode
}

\begin{document}

\title{
Missing Mass Matters: Neutrinophilic Scalar in $D$-Meson Decays
}

\author{Jin-Man Cai}
\email{caijm7@mail2.sysu.edu.cn}
\affiliation{School of Physics and Astronomy, Sun Yat-sen University, Zhuhai 519082, China}
\author{Gang Li}
\email{ligang65@mail.sysu.edu.cn}
\affiliation{School of Physics and Astronomy, Sun Yat-sen University, Zhuhai 519082, China}
\affiliation{
Guangdong Provincial Key Laboratory of Quantum Metrology and Sensing, Sun Yat-Sen University, Zhuhai 519082, China}

\date{\today}

\begin{abstract}

A light neutrinophilic scalar $\phi$, which couples predominantly to neutrinos, can be probed through leptonic meson decays for scalar masses $10~\mathrm{MeV}\lesssim m_\phi\lesssim\mathcal{O}(100)~\mathrm{MeV}$. We study neutrinophilic-scalar emission in $D$-meson decays at BESIII
and the future Super Tau-Charm Facility (STCF), focusing on $D_{(s)}^+\to e^+\bar{\nu}_\alpha\phi$. Accounting for the $m_\phi$-dependent missing-mass squared ($M_{\rm miss}^2$) distribution within the experimentally selected region, we find the resulting $D^+$ constraint to be substantially weaker than previous estimates based on the integrated three-body decay rate. Extending the $M_{\rm miss}^2$ coverage provides sensitivity to heavier scalars. With the full kinematic range of $M_{\rm miss}^2$ and an integrated luminosity of $1~\mathrm{ab}^{-1}$, the projected STCF sensitivities surpass the existing kaon bound for $m_\phi\gtrsim380~\mathrm{MeV}$ in the $D^+$ decay and $m_\phi\gtrsim310~\mathrm{MeV}$ in the $D_s^+$ decay.
We further discuss direct searches for $D_{(s)}^+\to\mu^+\bar{\nu}_\alpha\phi$ and their potential for probing the neutrinophilic scalar. A complete analytic expression for the three-body decay width is also derived with full dependence on the scalar and charged-lepton masses, including its behavior in the $m_\phi\to0$ and $m_\ell\to0$ limits.

\end{abstract}

\maketitle

\section{Introduction}
\label{sec:intro}

Neutrino oscillation experiments have established that neutrinos are
massive, providing compelling evidence for physics beyond the Standard
Model (SM). Extensions that address the origin of neutrino masses can
also give rise to new neutrino interactions. In addition to
non-standard neutrino interactions with matter, neutrino
self-interactions
have also received increasing
attention~\cite{Berryman:2022hds}.
Such interactions can be mediated by
a light (pseudo)scalar mediator $\phi$, which may arise
naturally in Majoron models~\cite{Chikashige:1980ui,Schechter:1981cv,Gelmini:1980re} or be embedded in other ultraviolet (UV)-complete frameworks~\cite{Berryman:2018ogk,Kelly:2019wow,deGouvea:2019qaz,Dev:2021axj}. Since $\phi$ couples
predominantly to neutrinos, it is commonly referred to as {\it neutrinophilic}. Importantly, its effective couplings to neutrinos need not be fixed solely by the observed neutrino masses, allowing them to be phenomenologically sizable~\cite{Berryman:2018ogk,Dev:2021axj,Berryman:2022hds}.

A neutrinophilic scalar $\phi$ can be probed through rare meson,
charged-lepton, and gauge-boson decays~\cite{Barger:1981vd,Gelmini:1982rr,Lessa:2007up,Pasquini:2015fjv,
Berryman:2018ogk,Brdar:2020nbj,Dev:2024ygx,Dev:2025tdv,Zhang:2026yez},
precision electroweak observables~\cite{Zhang:2024meg,Foroughi-Abari:2026ofc},
neutrino beamstrahlung~\cite{Berryman:2018ogk,Kelly:2021mcd,Liu:2024ywd,Adhikary:2024tvl},
neutrino scattering~\cite{Foroughi-Abari:2025mhj},
and collider searches with missing energy~\cite{deGouvea:2019qaz,Dev:2021axj,Liu:2024ywd}.
Leptonic decays of charged pseudoscalar mesons $P^\pm$ provide strong
laboratory constraints for
$10~\mathrm{MeV}\lesssim m_\phi\lesssim\mathcal{O}(100)~\mathrm{MeV}$~\cite{Pasquini:2015fjv,Berryman:2018ogk,Dev:2024ygx}. While the SM
two-body decay $P^+\to\ell^+\nu_\ell$ is helicity suppressed, emission
of $\phi$ from the neutrino line opens the three-body decay channel
$P^+\to\ell^+\bar{\nu}_\alpha\phi$ without the same
suppression~\cite{Berryman:2018ogk,deGouvea:2019qaz,Dev:2024ygx,deLima:2026fsy,Altmannshofer:2026opc}.
The couplings of $\phi$ to neutrinos can therefore be constrained using
both integrated decay rates and differential distributions. For light mesons, searches for additional invisible
particles and spectral distortions in $\pi^+$ and $K^+$ decays constrain
the neutrino-scalar couplings within the kinematically accessible
mass region $m_\phi<m_P-m_\ell$~\cite{PIENU:2021clt,NA62:2021bji,
Pasquini:2015fjv,Dev:2024ygx,deLima:2026fsy,Altmannshofer:2026opc}.

As $m_\phi$ approaches the kinematic limit of kaon decays, the larger
available phase space in $D^+$ and $D_s^+$ decays extends the
sensitivity to higher $m_\phi$. The SM leptonic decays of $D^+$ and $D_s^+$ have been measured by
BESIII~\cite{BESIII:2013iro,BESIII:2024kvt,
AblikimMai:2025heh,BESIII:2023cym} and Belle~\cite{Belle:2013isi}. Such measurements have been used to
constrain neutrino-scalar interactions by comparing the integrated
three-body decay rate with experimental limits or
uncertainties~\cite{Berryman:2018ogk,deGouvea:2019qaz,Dev:2025tdv}. However, the
continuous, $m_\phi$-dependent kinematic distribution of
$D_{(s)}^+\to\ell^+\bar{\nu}_\alpha\phi$ renders the resulting
constraints sensitive to selected kinematic regions~\cite{Pasquini:2015fjv}. The recent BESIII search for
$D^+\to e^+\nu_e$, for example, extracts the signal yield from a fit
to the missing-mass-squared $M_{\rm miss}^2$ distribution with
$M_{\rm miss}^2<0.25~\mathrm{GeV}^2$~\cite{AblikimMai:2025heh}. Since the fraction of the three-body signal falling within this region
depends on $m_\phi$, its $M_{\rm miss}^2$ distribution and the restricted
experimental coverage must be treated consistently when deriving
constraints on the neutrino-scalar coupling.

In this work, we investigate neutrinophilic-scalar emission in
$D^+$ and $D_s^+$ decays across the accessible scalar-mass range and focus on the electron channel, where the SM leptonic decay is strongly helicity suppressed. We derive the
differential decay rate and a complete analytic expression for
$P^+\to\ell^+\bar{\nu}_\alpha\phi$, retaining the full
dependence on the scalar and charged-lepton masses.
Using the recent BESIII search for $D^+\to e^+\nu_e$ based on
$20.3~\mathrm{fb}^{-1}$ of data collected at
$\sqrt{s}=3.773~\mathrm{GeV}$~\cite{AblikimMai:2025heh},
we reinterpret this search to constrain
$D^+\to e^+\bar{\nu}_\alpha\phi$ by accounting for the
$m_\phi$-dependent signal fraction within the selected
$M_{\rm miss}^2$ region. The resulting constraint is substantially
weaker than previous estimates based on the integrated three-body decay rate
and remains below the existing kaon bound. We further explore the
potential of direct searches for $D_{(s)}^+\to e^+\bar{\nu}_\alpha\phi$
over a broader $M_{\rm miss}^2$ range and project their reach at STCF
using the luminosity benchmarks of its conceptual
design~\cite{Achasov:2023gey}.
As a benchmark, we consider the full kinematic range of
$M_{\rm miss}^2$ and estimate the corresponding backgrounds. With an
integrated luminosity of $1~\mathrm{ab}^{-1}$ at STCF, the projected
$D^+$ sensitivity surpasses the kaon bound for
$m_\phi\gtrsim380~\mathrm{MeV}$. The $D_s^+$ sensitivity is stronger
than that of $D^+$ throughout their common scalar-mass range and
surpasses the kaon bound for $m_\phi\gtrsim310~\mathrm{MeV}$.

The remainder of this paper is organized as follows. In Sec.~\ref{sec:model}, we introduce the neutrino-scalar interaction
and derive the differential decay rate, with particular emphasis on
the missing-mass distribution and the experimentally selected region. In
Sec.~\ref{sec:analysis}, we present the current constraints and future
sensitivities of $D^+$ and $D_s^+$ decays and discuss the prospects
for the muon channels. Section~\ref{sec:conclusion} summarizes our main
results. Details of the derivation of the analytic three-body decay
width, including its behavior in the $m_\phi\to0$ and $m_\ell\to0$ limits, are provided in the
appendix.

\section{Neutrino-Scalar Interactions in Meson Decays}
\label{sec:model}

We consider the interaction of a light scalar $\phi$ with neutrinos~\cite{Zhang:2024meg,Brdar:2020nbj,Zhang:2026yez,deLima:2026fsy}
\begin{align}
\mathcal{L}_{\rm int}
=
\frac{1}{2} g_{\alpha\beta}
\overline{\nu_\alpha^c} \nu_\beta \phi
+\mathrm{h.c.}\,,
\label{eq:nu_scalar_interaction}
\end{align}
where $\nu_\alpha$ denotes a left-handed neutrino field and
$\nu_\alpha^c$ its charge conjugate. The flavor indices are
$\alpha,\beta=e,\mu,\tau$, and the coupling matrix is symmetric,
$g_{\alpha\beta}=g_{\beta\alpha}$.

For a charged pseudoscalar meson $P^+$, the SM leptonic decay
$P^+\to\ell^+\nu_\ell$ has the tree-level width~\cite{Rosner:2015wva}
\begin{align}
\Gamma_{\rm SM}
(P^+\to\ell^+\nu_\ell)
&=
\frac{G_F^2 f_P^2 |V_{qq'}|^2 m_P m_\ell^2}{8\pi}
(1-\alpha)^2\,,
\label{eq:SM_width_massless_nu}
\end{align}
where the neutrino masses are neglected. Here, $\alpha\equiv m_\ell^2/m_P^2$ (not to be confused with the
neutrino-flavor index), $G_F$ is the Fermi constant,
$f_P$ is the meson decay constant, $V_{qq'}$ is the relevant element
of the Cabibbo--Kobayashi--Maskawa matrix, and $m_P$ and
$m_\ell$ denote the meson and charged-lepton masses, respectively.
The factor of $m_\ell^2$ reflects the helicity suppression
associated with 
the SM charged-current
interaction.

The interaction in Eq.~\eqref{eq:nu_scalar_interaction} gives rise to
scalar emission from the final-state neutrino, $P^+ \to \ell^+ \bar{\nu}_\alpha\phi$, if it is kinematically allowed.
Denoting the total four-momentum of the invisible particles by
$q=p_\nu+p_\phi$, we introduce the dimensionless variables
$x=q^2/m_P^2$ and $\gamma=m_\phi^2/m_P^2$, with
$\gamma\leq x\leq(1-\sqrt{\alpha})^2$ for negligible neutrino masses.

The differential decay rate can be written as~\cite{Pasquini:2015fjv,Barger:1981vd}
\begin{align}
\frac{\dd\Gamma_{\rm BSM}}{\dd x}
&=
\Gamma(P^+\to\ell^+\nu_\ell^\ast)
\frac{|g_{\ell\alpha}|^2}{32\pi^2}
\frac{(x-\gamma)^2}{x^3}\,,
\label{eq:dGamma_dx}
\end{align}
where the effective two-body decay width involving a virtual neutrino
$\nu_\ell^\ast$ is given by~\cite{Pasquini:2015fjv,Barger:1981vd,Shrock:1980ct}
\begin{align}
\Gamma(P^+\to\ell^+\nu_\ell^\ast)
&=
\frac{G_F^2 f_P^2 |V_{qq'}|^2m_P^3}{8\pi}
\lambda^{1/2}(1,x,\alpha)
\nonumber\\
&\quad\times
\left[
x+\alpha-(x-\alpha)^2
\right]\,,
\label{eq:Gamma_virtual_nu}
\end{align}
with the K\"all\'en function being defined as
\begin{align}
\lambda(a,b,c)
&=
a^2+b^2+c^2-2ab-2ac-2bc\,.
\label{eq:Kallen_function}
\end{align}

In the limit $m_\ell\to0$, Eq.~\eqref{eq:dGamma_dx} reduces to
\begin{align}
\left.
\frac{\dd\Gamma_{\rm BSM}}{\dd x}
\right|_{\alpha=0}
&=
\frac{G_F^2 f_P^2 |V_{qq'}|^2m_P^3
|g_{\ell\alpha}|^2}{256\pi^3}
\nonumber\\
&\quad\times
\frac{(1-x)^2(x-\gamma)^2}{x^2}\,.
\label{eq:dGamma_dx_massless_lepton}
\end{align}
The three-body decay width therefore remains nonzero as $m_\ell\to0$, in
contrast to the SM two-body decay width in
Eq.~\eqref{eq:SM_width_massless_nu}, and
therefore avoids the helicity suppression. 
The three-body decay width is
obtained by integrating $\dd\Gamma_{\rm BSM}/\dd x$ over the full
kinematic range $\gamma\leq x\leq(1-\sqrt{\alpha})^2$.

Experimentally, the analyses cover only part of the kinematically
allowed region. The corresponding signal rate is therefore determined
by the differential spectrum within the selected region rather than by
the full three-body decay width. For leptonic meson decays, the invariant mass squared of the invisible particles corresponds to the missing-mass squared,
\begin{align}
   M_{\rm miss}^2\equiv q^2=xm_P^2\,.
   \label{eq:MM2}
\end{align}
The SM two-body decay is concentrated around
$M_{\rm miss}^2\simeq0$, up to detector resolution, whereas the
three-body signal has a continuous distribution beginning at
$M_{\rm miss}^2=m_\phi^2$. The corresponding differential decay rate is
\begin{align}
\frac{\dd\Gamma_{\rm BSM}}
     {\dd M_{\rm miss}^2}
&=
\frac{1}{m_P^2}
\frac{\dd\Gamma_{\rm BSM}}{\dd x}\,.
\label{eq:dGamma_dMmiss2}
\end{align}
Such an observable has been used in analyzing the leptonic decays of 
$K^+$~\cite{NA62:2021bji,Krnjaic:2019rsv}, $D^+$~\cite{BESIII:2013iro,
AblikimMai:2025heh}, and $D_s^+$~\cite{
Belle:2013isi,BESIII:2023cym} mesons.

For a selected window of $M_{\rm miss}^2$, the corresponding
three-body decay width is
\begin{align}
\Gamma_{\rm BSM}^{\rm sel}
&=
\int_{x_{\min}}^{x_{\max}}
\dd x\,
\frac{\dd\Gamma_{\rm BSM}}{\dd x}\,,
\label{eq:Gamma_BSM_window}
\end{align}
where $x_{\min}$ and $x_{\max}$ are determined by the overlap between
the full kinematic range
$\gamma\leq x\leq(1-\sqrt{\alpha})^2$
and the experimentally selected $M_{\rm miss}^2$ range expressed in
terms of $x$.

\begin{figure}[!htbp]
\centering
\includegraphics[width=0.35\textwidth]{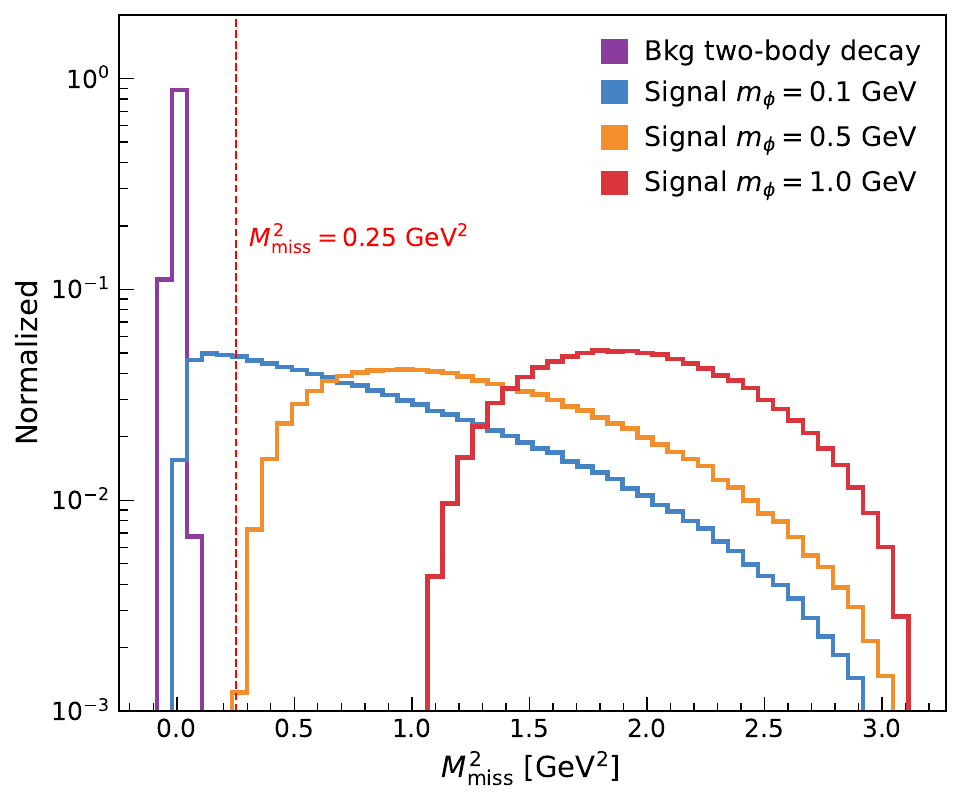}
\caption{Missing-mass squared $M_{\rm miss}^2$ distributions for
$D^+\to e^+\bar{\nu}_\alpha\phi$,
assuming the momentum resolution of electrons
$\sigma_{p_T}/p_T < 0.5\%$~\cite{AblikimMai:2025heh,Achasov:2023gey}.
The
blue, orange, and red curves correspond to the signals with $m_\phi=0.1$, $0.5$, and
$1~\mathrm{GeV}$, respectively. The purple curve denotes the SM $D^+\to e^+\nu_e$ background, and the red dashed line marks the upper boundary of the selected window,
$M_{\rm miss}^2=0.25~\mathrm{GeV}^2$.}
\label{fig:distribution}
\end{figure}

The fraction of the three-body signal in the selected kinematic
region varies with $m_\phi$. As $m_\phi$ increases, the signal
distribution shifts toward larger $M_{\rm miss}^2$, while the available three-body phase space decreases, as illustrated in
Fig.~\ref{fig:distribution}. Consequently, the $m_\phi$-dependent
signal fraction within the experimentally selected region must be
incorporated when deriving constraints on the neutrino-scalar
coupling.

We note that the $m_\phi$ dependence can also be expressed in terms of
the charged-lepton energy $E_\ell$ in the rest frame of the decaying
meson,
\begin{align}
E_\ell
&=
\frac{m_P^2+m_\ell^2-q^2}{2m_P}
=
\frac{m_P}{2}(1+\alpha-x)\,.
\label{eq:El_x_relation}
\end{align}
This implies that a larger $M_{\rm miss}^2$ corresponds to a lower charged-lepton
energy $E_\ell$. The charged-lepton energy spectrum can therefore
provide a complementary probe of the neutrinophilic scalar, as in pion
decays~\cite{PIENU:2021clt,Altmannshofer:2026opc,deLima:2026fsy}.\footnote{Alternatively, the charged-lepton momentum spectrum can be
used for processes with a muon and invisible particles in the final
state~\cite{PIENU:2020las,PIENU:2021clt,E949:2016gnh}.}
For the collider processes considered in Sec.~\ref{sec:analysis},
however, the Lorentz-invariant variable $M_{\rm miss}^2$ is
particularly convenient, as it can be constructed directly without transforming to the rest frame of
the decaying meson.

Before closing this section, we comment on the infrared (IR) behavior of
the three-body decay width. In the limit $m_\phi\to0$, the three-body
decay width develops an IR divergence, which is canceled by the
virtual correction to the corresponding two-body decay width. For the electron channel
considered in this work, the finite virtual contribution is strongly
suppressed by $m_e^2/m_P^2$ and can therefore be safely neglected in
our numerical analysis. The complete analytic expression for the three-body decay width and
its IR behavior are discussed in
Appendix~\ref{app:calculation_details}.

\section{\tf{$D$}{D}-Meson Decays at BESIII and STCF}
\label{sec:analysis}

We begin with the $D^+$ decay at BESIII, where $D^+D^-$ pairs are
produced through $e^+e^-\to\psi(3770)\to D^+D^-$ at
$\sqrt{s}=3.773~\mathrm{GeV}$. The current data sample corresponds to
an integrated luminosity of $20.3~\mathrm{fb}^{-1}$~\cite{BESIII:2024kvt,AblikimMai:2025heh}. We focus on the electron channel, with one $D$ meson fully
reconstructed as the tag and an electron accompanied by missing energy
on the signal side.\footnote{Charge-conjugate modes are implicitly included throughout.}

The electron channel is particularly sensitive, since the SM leptonic
decay is strongly suppressed. Charged-pion backgrounds can be
efficiently reduced by electron--pion
identification~\cite{Qin:2008pid}. The remaining backgrounds mainly arise from semileptonic $D$ decays
with an $e^\pm$~\cite{AblikimMai:2025heh}.

From the $M_{\rm miss}^2$ distribution reported by BESIII~\cite{AblikimMai:2025heh}, we take $M_{\rm miss}^2=0.25~{\rm GeV}^2$ as the upper boundary,
corresponding to $x_{\max}\simeq 0.07$. For a tagged $D$ sample, the expected
number of $D^+\to e^+ \bar{\nu}_\alpha\phi$ signal events in the selected region is expressed as
$N_{\rm sig}=N_{\rm tag}\epsilon_{\rm sig}{\rm Br}_{\rm BSM}^{\rm sel}$,
where $N_{\rm tag}$ denotes the number of reconstructed tag-side $D$ mesons
and $\epsilon_{\rm sig}$ is the signal efficiency. 
The BESIII sample
contains $N_{\rm tag}\simeq1.07\times10^7$ reconstructed tags.
We adopt $\epsilon_{\rm sig}=0.67$ as the benchmark signal efficiency
based on the $D^+\to e^+\nu_e$ search~\cite{AblikimMai:2025heh}.
Here ${\rm Br}_{\rm BSM}^{\rm sel}
=\Gamma_{\rm BSM}^{\rm sel}/\Gamma_{D^+}$,
where $\Gamma_{D^+}$ is the total decay width of $D^+$
and $\Gamma_{\rm BSM}^{\rm sel}$ is given in Eq.~\eqref{eq:Gamma_BSM_window}.

\begin{figure}[!htbp]
\centering
\includegraphics[width=0.4\textwidth]{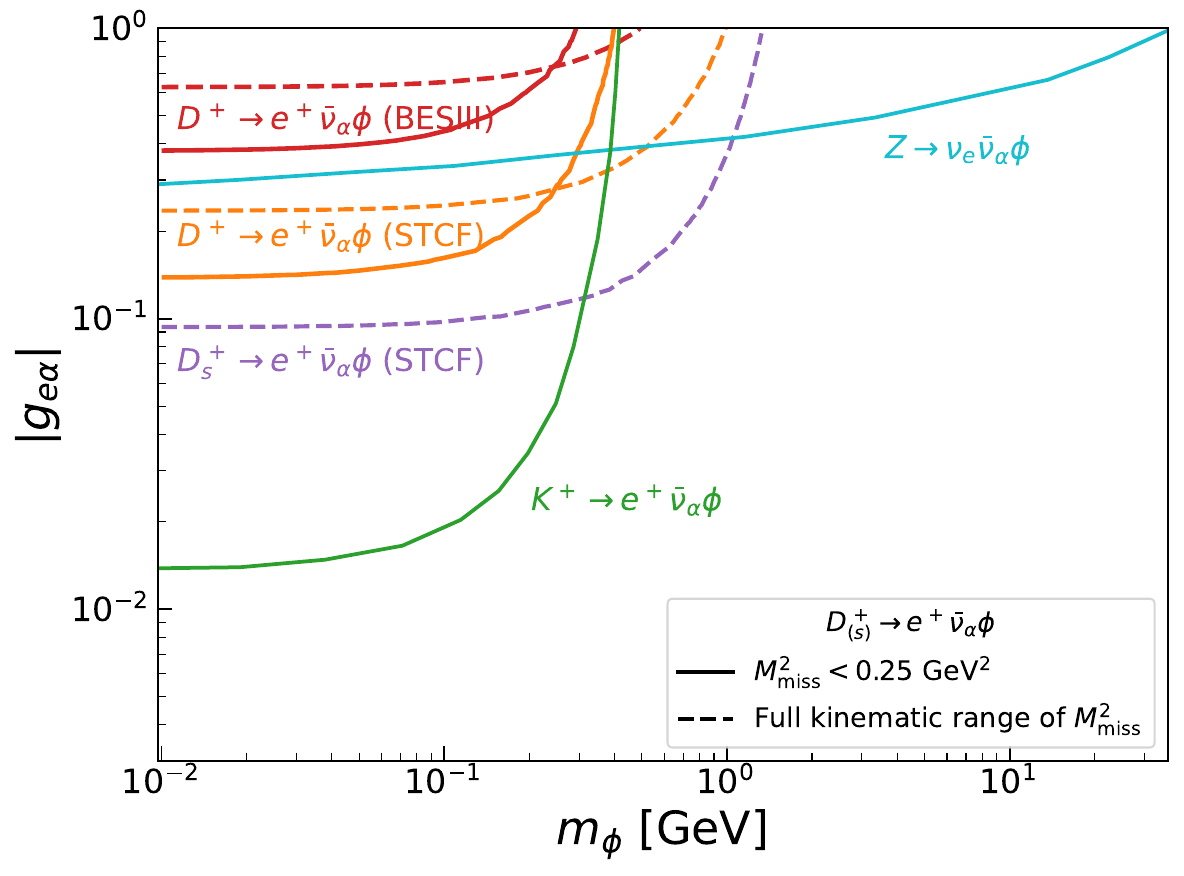}
\caption{Current constraints and projected sensitivities at 90\% C.L. for the neutrino-scalar coupling $|g_{e\alpha}|$ ($\alpha=e,\mu$, or $\tau$)
as a function of the neutrinophilic-scalar mass $m_\phi$.
The red curves show the sensitivities based on the current BESIII
data sample, while the orange curves show the STCF projections for
$D^+$ decays. The solid curves correspond to
$M_{\rm miss}^2<0.25~\mathrm{GeV}^2$, while the dashed curves are
based on the full kinematic range of $M_{\rm miss}^2$.
The purple dashed curve shows the corresponding STCF projection for
the $D_s^+$ decay.
The blue and green curves denote the existing bounds from $Z$-boson
and kaon decays, respectively~\cite{Berryman:2018ogk,Brdar:2020nbj}.
}
\label{fig:De_channel}
\end{figure}

For the current BESIII selection with
$M_{\rm miss}^2<0.25~\mathrm{GeV}^2$, we take
$N_{\rm bkg}= 249$ from the fitted background
distribution~\cite{AblikimMai:2025heh}.
We then estimate the sensitivity at 90\% confidence level (C.L.) using the approximate criterion
$N_{\rm sig}/\sqrt{N_{\rm sig}+N_{\rm bkg}}=1.64$. This yields the BESIII
constraint shown by the red solid curve in Fig.~\ref{fig:De_channel}.
Since the flavor of the missing neutrino cannot be identified
experimentally, we consider one neutrino flavor at a time, for which
the same sensitivity applies to $|g_{ee}|$, $|g_{e\mu}|$, and
$|g_{e\tau}|$. Flavor-changing couplings are subject to additional
constraints from other processes; see, e.g.,
Ref.~\cite{Zhang:2026yez} and references therein.

Previous studies derived $D$-meson constraints that can exceed the kaon
bound at larger $m_\phi$ by applying experimental branching-ratio
limits to the integrated three-body decay
rate~\cite{Berryman:2018ogk,deGouvea:2019qaz,Dev:2025tdv}.
Accounting instead for the $m_\phi$-dependent fraction of the signal
within the experimentally probed $M_{\rm miss}^2$ region weakens the
sensitivity, and the BESIII constraint from $D^+\to e^+\bar{\nu}_\alpha\phi$ remains below the
kaon bound, as shown in Fig.~\ref{fig:De_channel}.

For the STCF projection, we consider $D^+D^-$ production at
$\sqrt{s}=3.773~\mathrm{GeV}$ with an integrated luminosity of
$1~\mathrm{ab}^{-1}$~\cite{Achasov:2023gey}. We obtain $N_{\rm tag}$ and $N_{\rm bkg}$ by scaling their BESIII values by the ratio of integrated luminosities.
The same $M_{\rm miss}^2$ region, signal efficiency, and statistical
procedure as for BESIII are adopted. The resulting projected
sensitivity is shown by the orange solid curve in
Fig.~\ref{fig:De_channel}.

Since the signal shifts toward larger $M_{\rm miss}^2$ as $m_\phi$
increases, the currently selected $M_{\rm miss}^2$ window limits the reach
for heavier $\phi$. To explore the potential sensitivity beyond this
region, we consider the full kinematic range of
$M_{\rm miss}^2$, corresponding to $x\lesssim1$, as a benchmark and
estimate the backgrounds as described below.

The dominant backgrounds
are $D^+\to\pi^0e^+\nu_e$ and $D^+\to K_L^0e^+\nu_e$, with branching ratios of about $3.7\times10^{-3}$ and $4.5\times10^{-2}$,
respectively~\cite{ParticleDataGroup:2026mpi,BESIII:2015jmz}. These backgrounds have
been explicitly considered in a recent BESIII analysis, where
deep-learning method reduced the overall background by more than
two orders of magnitude~\cite{BESIII:2025mnc}. We thus conservatively retain
10\% of these modes in estimating the background. Other semileptonic
backgrounds, including $D^+\to\eta e^+\nu_e$, $\rho^0e^+\nu_e$, and
$\omega e^+\nu_e$, have branching ratios at the $10^{-3}$
level~\cite{ParticleDataGroup:2026mpi}. 
Since these modes contain additional visible particles that can be
vetoed, as implemented in the BESIII
analysis of $D^+\to e^+\nu_e$~\cite{AblikimMai:2025heh}, we retain 1\% of their contributions in the
background estimate.
These assumptions give a benchmark background yield of
$N_{\rm bkg}\simeq5.2\times10^4$ events at BESIII over the full $M_{\rm miss}^2$ range, which is scaled by the ratio of integrated
luminosities for the corresponding STCF projection.
A dedicated experimental analysis would be
required to determine the optimal selection and the corresponding
background rejection. 

The resulting BESIII and STCF sensitivities over the full
$M_{\rm miss}^2$ range are shown by the red and orange dashed curves in
Fig.~\ref{fig:De_channel}, respectively.
Extending the
$M_{\rm miss}^2$ coverage increases the background and thus weakens
the sensitivity at lower $m_\phi$, while improving the sensitivity at
larger $m_\phi$
and allows the projected $D^+$ sensitivity at STCF to surpass the
existing kaon bound~\cite{Berryman:2018ogk}\,\footnote{This bound was derived in Ref.~\cite{Berryman:2018ogk} by
reinterpreting the experimental upper limit on the branching ratio
$\mathrm{Br}(K^+\to e^+\nu_e\nu\bar{\nu})<6\times10^{-5}$~\cite{ParticleDataGroup:2026mpi} as the 90\% C.L. constraint on the branching
ratio of $K^+\to e^+\bar{\nu}_\alpha\phi$.
In Ref.~\cite{Dev:2024ygx}, instead, the bound was obtained by
requiring the $K^+\to e^+\bar{\nu}_\alpha\phi$ contribution to lie
within the $1.64\sigma$ uncertainty of the
$K^+\to e^+\nu_e$ partial width. We use the former since the corresponding
decay more closely resembles the kinematics of the three-body signal
considered here. The SM two-body signal relevant to the latter is
concentrated near $M_{\rm miss}^2\simeq0$, while the three-body signal
has a continuous $M_{\rm miss}^2$ distribution similar to that shown
in Fig.~\ref{fig:distribution}.} for
$m_\phi\gtrsim 380~\mathrm{MeV}$. The difference between the solid and dashed curves demonstrates the
dependence of the sensitivity on the kinematic selection, motivating
a dedicated optimization of the $M_{\rm miss}^2$ region.

The $D_s^+\to e^+\bar{\nu}_\alpha\phi$ channel is also considered, for
which the larger $D_s^+$ mass enables sensitivity to heavier $\phi$. At $\sqrt{s}=4.009~\mathrm{GeV}$, we consider an integrated luminosity
of $1~\mathrm{ab}^{-1}$ at STCF, corresponding to one year of data
taking~\cite{Achasov:2023gey,Liu:2021qio}, and take
$N_{\rm tag}\simeq3.45\times10^7$ reconstructed $D_s^-$ tags based on
the STCF simulation in Ref.~\cite{Liu:2021qio}. We adopt the same
benchmark signal efficiency, $\epsilon_{\rm sig}=0.67$, as for the
$D^+$ analysis, given their similar masses and the same charged-lepton
final state.

For the $e^++E_{\rm miss}$ backgrounds, we use the STCF simulation of $D_s^+\to\tau^+\nu_\tau$, followed by $\tau^+\to e^+\nu_e\bar{\nu}_\tau$, as a reference~\cite{Li:2021ala}.
This SM process and
$D_s^+\to K_L^0e^+\nu_e$ constitute the main backgrounds, with
branching ratios of about $9.6\times10^{-3}$ and
$1.4\times10^{-3}$, respectively~\cite{ParticleDataGroup:2026mpi,
BESIII:2018xre}.
A conservative retention factor of 10\% is assumed for these modes in estimating the background.
Other semileptonic backgrounds include
$D_s^+\to\eta e^+\nu_e$, $\eta'e^+\nu_e$, and
$\phi(1020)e^+\nu_e$, with branching ratios of about 2\%, 1\%, and
2\%, respectively~\cite{ParticleDataGroup:2026mpi}, while
$D_s^+\to K^{*0}e^+\nu_e$ has a branching ratio at the
$10^{-3}$ level~\cite{BESIII:2018xre}. Given the additional visible
particles and the use of extra neutral energy and event topology for
background rejection~\cite{Li:2021ala}, a retention factor of 1\% is
assumed for these modes.
These assumptions give an expected background yield of
$N_{\rm bkg}\simeq5.7\times10^4$ events at STCF
over the full $M_{\rm miss}^2$ range.

We then estimate the $D_s^+$ sensitivity over the full kinematic range
of $M_{\rm miss}^2$. As shown in Fig.~\ref{fig:De_channel}, the
projected $D_s^+$ sensitivity at STCF is stronger than that for $D^+$
throughout their common scalar-mass range and surpasses the existing kaon
bound for $m_\phi\gtrsim310~\mathrm{MeV}$. Compared with the $D^+$ decay, the larger available phase space in
the $D_s^+$ decay provides sensitivity to heavier neutrinophilic
scalars. Despite the smaller number of $D_s$ tags, the enhancement
from $|V_{cs}/V_{cd}|^2\sim19$ and the lower background lead to
stronger sensitivity.

For completeness, we assess the sensitivity of the existing Belle data
collected around the $\Upsilon(4S)$ and $\Upsilon(5S)$
resonances~\cite{Belle:2013isi}. The signal-side $D_s$ is inclusively
reconstructed in recoil against a tag-side charm hadron, fragmentation
particles, and the photon from $D_s^*\to D_s\gamma$. A search for $D_s^+\to e^+\bar{\nu}_\alpha\phi$ could then use $M_{\rm miss}^2(D_{\rm tag}K_{\rm frag}X_{\rm frag}\gamma e)$, which corresponds to the $M_{\rm miss}^2$ variable defined in
Eq.~\eqref{eq:MM2}. Using the Belle sample, with approximately $9.4\times10^4$ inclusively reconstructed $D_s$ mesons~\cite{Belle:2013isi}, we estimate that the resulting constraint is weaker than that from BESIII and therefore do not show it in Fig.~\ref{fig:De_channel}.

The muon channels in $D$-meson decays also provide sensitivity to neutrino-scalar
interactions, with backgrounds distinct from those in the electron
channel. For $D^+\to\mu^+\bar{\nu}_\alpha\phi$, the SM
$D^+\to\mu^+\nu_\mu$ decay is less helicity suppressed, and
pion--muon misidentification leads to sizable hadronic backgrounds.
In particular, $D^+\to K_L^0\pi^+$ peaks around
$M_{\rm miss}^2\simeq0.25\gev^2$~\cite{BESIII:2013iro}.
A quantitative sensitivity estimate therefore requires a dedicated
analysis of the $m_\phi$-dependent signal distribution and backgrounds.

For $D_s^+\to\mu^+\bar{\nu}_\alpha\phi$, additional hadronic
backgrounds become relevant when considering a broader
$M_{\rm miss}^2$ range~\cite{BESIII:2023cym,Liu:2021qio}.
Analogous to $D^+\to K_L^0\pi^+$~\cite{BESIII:2013iro},
$D_s^+\to K_L^0K^+$ constitutes a potential background.
Despite its sizable branching fraction, $K^+$ is much less likely to
be misidentified as $\mu^+$ than $\pi^+$, and the background can be
further reduced through kaon identification.
Another relevant hadronic mode is $D_s^+\to\phi(1020)\pi^+$, for which
the subsequent $\phi(1020)$ decay produces additional visible
particles that can be used for background suppression.
These features facilitate a direct search
for $D_s^+\to\mu^+\bar{\nu}_\alpha\phi$ at STCF.

\section{Conclusion}
\label{sec:conclusion}

In this work, we have studied neutrinophilic-scalar emission in
$D$-meson decays, focusing on
$D_{(s)}^+\to\ell^+ \bar{\nu}_\alpha\phi$ at BESIII and STCF.
For $D^+\to e^+ \bar{\nu}_\alpha\phi$, we reinterpret the latest
$20.3~\mathrm{fb}^{-1}$ BESIII search at
$\sqrt{s}=3.773~\mathrm{GeV}$ by accounting for the continuous,
$m_\phi$-dependent $M_{\rm miss}^2$ distribution of the three-body
signal. The current search probes only
$M_{\rm miss}^2<0.25~\mathrm{GeV}^2$, corresponding to
$x\lesssim0.07$, and therefore captures only part of the signal,
particularly for larger $m_\phi$. Accounting for the signal fraction
within this region substantially weakens the constraint relative to
estimates based on the fully integrated three-body decay rate, leaving the
current BESIII bound weaker than the existing kaon bound.

The limited $M_{\rm miss}^2$ coverage of the current search also
motivates a dedicated analysis of the existing BESIII data over a
wider kinematic range. To assess the potential reach at larger
$m_\phi$, we consider the full kinematic range of $M_{\rm miss}^2$ and estimate the corresponding backgrounds in our analysis. Although the increased background weakens
the coupling constraint at lower $m_\phi$, the broader coverage
provides sensitivity to the signal at higher $M_{\rm miss}^2$ and
thereby extends the reach to heavier scalars. A dedicated experimental
analysis would be required to determine the signal efficiency and
backgrounds over the enlarged region. For STCF, we consider both the restricted and full
$M_{\rm miss}^2$ ranges, with the corresponding background estimates.
With the full kinematic range and an integrated luminosity of
$1~\mathrm{ab}^{-1}$, the projected
$D^+\to e^+ \bar{\nu}_\alpha\phi$ sensitivity surpasses the kaon bound
for $m_\phi\gtrsim380~\mathrm{MeV}$. The projected $D_s^+$ sensitivity is stronger than that of $D^+$
throughout their common scalar-mass range and surpasses the kaon
bound for $m_\phi\gtrsim310~\mathrm{MeV}$.

These results demonstrate that the sensitivity of $D$-meson decays to
neutrinophilic scalars depends crucially on the experimentally probed
$M_{\rm miss}^2$ range. While the coverage of the current
BESIII search limits its sensitivity, considering the full kinematic
range as a benchmark allows the larger phase space of $D^+$ and
$D_s^+$ decays to probe heavier scalars beyond the reach of existing
kaon constraints.

\section{Acknowledgments}

This work is supported by the National Natural Science Foundation of China under Grants No.~12347105 and No.~12505127, and the Guangdong Basic and Applied Basic Research Foundation (2024A1515012668).

\appendix
\counterwithin{figure}{section}
\counterwithin{table}{section}

\section{Analytic three-body decay width}
\label{app:calculation_details}

From Eqs.~\eqref{eq:dGamma_dx} and \eqref{eq:Gamma_virtual_nu}, the three-body decay width can be written as
\begin{align}
\Gamma(P^+\to\ell^+\bar{\nu}_\alpha\phi)
&=
\frac{G_F^2 f_P^2 |V_{qq'}|^2m_P^3 |g_{\ell\alpha}|^2}{256\pi^3}I(\alpha, \gamma)\,,
\label{eq:Gamma_tot}
\end{align}
where the dimensionless phase-space integral $I(\alpha,\gamma)$ is defined as
\begin{align}
I(\alpha, \gamma)=\int_{\gamma}^{(1-\sqrt{\alpha})^2}
&\left[
x+\alpha-(x-\alpha)^2
\right]\,\nn\\
&\times \lambda^{1/2}(1,x,\alpha) 
\frac{(x-\gamma)^2}{x^3} \dd x\,.
\end{align}

The K\"all\'en function can be factorized as
\begin{align}
\lambda(1,x,\alpha)
&=(1+\alpha-x)^2-4\alpha\,.
\end{align}
We then introduce the substitution
\begin{align}
x=1+\alpha-2\sqrt{\alpha}\cosh u\,,
\end{align}
which gives
\begin{align}
\lambda^{1/2}(1,x,\alpha)
&=2\sqrt{\alpha}\sinh u\,, \\
\dd x
&=-2\sqrt{\alpha}\sinh u\,\dd u\,.
\end{align}
Using standard hyperbolic-function identities,
the integral can be evaluated analytically, which yields
\begin{align}
I(\alpha, \gamma) 
&= (\alpha + 2\gamma) \left[ \frac{1 + \alpha - \gamma}{2}Q - 2\alpha U \right] + \frac{Q^3}{3}\nn \\
&\quad + A \left[ -(1 + \alpha)U - Q + 2(1 - \alpha)H \right]\nn \\
&\quad + B \left[ \frac{Q}{2\gamma^2} - \frac{(1 + \alpha)Q}{2(1 - \alpha)^2\gamma} - \frac{4\alpha}{(1 - \alpha)^3}H \right] \nn\\
&\quad + C \left[ U - \frac{2(1 + \alpha)}{1 - \alpha}H + \frac{Q}{\gamma} \right] \,.
\end{align}
Here, the auxiliary quantities are defined as
\begin{align}
A &= \alpha(1-\alpha)-\gamma^2-2(1+2\alpha)\gamma\,, \\
B &= \alpha(1-\alpha)\gamma^2\,, \\
C &= (1+2\alpha)\gamma^2-2\alpha(1-\alpha)\gamma\,, \\
Q &= \sqrt{(1+\alpha-\gamma)^2-4\alpha}\,, \\
U &= \ln\frac{1+\alpha-\gamma+Q}{2\sqrt{\alpha}}\,, \\
H &= \operatorname{artanh}
\sqrt{
\frac{(1+\sqrt{\alpha})^2[(1-\sqrt{\alpha})^2-\gamma]}
{(1-\sqrt{\alpha})^2[(1+\sqrt{\alpha})^2-\gamma]}
}\,.
\label{eq:auxiliary_quantities}
\end{align}
We have verified that the above analytic expression is in exact
agreement with direct numerical integration. A comparison with the complete analytic expression for the integrated
three-body decay width given in Ref.~\cite{Dev:2024ygx} reveals a
difference that increases with $m_\phi$.

\subsection{The \tf{$m_\phi \to 0$}{mphi -> 0} limit}
\label{sec:mphi-limit}

In the limit $m_\phi \to 0$, or equivalently $\gamma \to 0$, 
the auxiliary quantity $H$ can be expanded as
\begin{align}
H &= -\frac{1}{2}\ln\gamma
+ \frac{1}{2}\ln\frac{4}{\gamma^\prime}
+ \mathcal{O}(\gamma)\,, 
\end{align}
where 
\begin{align}
    \gamma^\prime \equiv \frac{1}{(1-\sqrt{\alpha})^2}-\frac{1}{(1+\sqrt{\alpha})^2}\,.
\end{align}
The integral in this limit then behaves as
\begin{align}
I(\alpha,\gamma)
& = 2\alpha(1-\alpha)^2(
-\frac{1}{2}\ln\gamma
+ \frac{1}{2}\ln\frac{4}{\gamma^\prime}) \nn \\
& \quad +\alpha(\frac{1-\alpha^2}{2}-2\alpha\ln\frac{1}{\sqrt{\alpha}})+\frac{(1-\alpha)^3}{3}\,.
\label{eq:I_beta_zero}
\end{align}
For finite $\alpha$, the term proportional to $\alpha\ln\gamma$
is IR divergent. This divergence is canceled by the corresponding divergence of the two-body virtual correction~\cite{Pasquini:2015fjv,Dev:2024ygx}, which is given by~\cite{deLima:2026fsy}
\begin{align}
\delta\Gamma(P^+ \to \ell^+\nu_\ell)
&=
-\frac{G_F^2 f_P^2 |V_{qq'}|^2 m_P^3 |g_{\ell\alpha}|^2}
{256\pi^3}
\,\alpha(1-\alpha)^2
\nn\\
&\quad\times
\left(
\ln\frac{\Lambda^2}{m_\phi^2}
+2\xi_{\rm UV}
\right)\,.
\label{eq:two-body-vir}
\end{align}
Here, $\Lambda$ is the UV cutoff scale and $\xi_{\rm UV}$ parametrizes
the finite UV-model dependence. We take
$\Lambda=500~{\rm GeV}$ and $\xi_{\rm UV}=1$, corresponding to the
heavy $SU(2)_L$-triplet completion as in Ref.~\cite{deLima:2026fsy}. After cancellation of the IR-divergent term by the corresponding
two-body virtual correction, the subsequent limit $\alpha\to0$ yields
$I(0,0)=1/3$.

\vspace{0.3cm}

\subsection{The \tf{$m_\ell \to 0$}{ml  -> 0} limit}

In the limit $m_\ell \to 0$, or equivalently $\alpha \to 0$, the auxiliary quantities $U$ and $H$ can be expanded as
\begin{align}
U &= -\frac{1}{2}\ln\alpha + \ln (1-\gamma) + \mathcal{O}(\alpha)\,, \\
H &= -\frac{1}{4}\ln\alpha
+ \frac{1}{2}\ln\frac{1-\gamma}{\gamma}
+ \mathcal{O}(\sqrt{\alpha})\,.
\end{align}
Both $U$ and $H$ are logarithmically divergent as $\alpha\to0$, but
these divergences cancel exactly after their coefficients are
combined, in contrast to the $m_\phi\to0$ limit at finite $\alpha$. The integral then reduces to
\begin{align}
I(0,\gamma)
=
\frac{1}{3}
+3\gamma
-3\gamma^2
-\frac{\gamma^3}{3}
+2\gamma(1+\gamma)\ln\gamma\,.
\label{eq:I_zero_gamma}
\end{align}
The resulting three-body decay width agrees with
Ref.~\cite{Berryman:2018ogk}. Since $\gamma\ln\gamma\to0$ as $\gamma\to0$, the above expression remains finite and also yields $I(0,0)=1/3$.

The two limiting procedures exhibit different IR structures.
For finite $\alpha$, the $m_\phi\to0$ limit of the real-emission
width contains an IR-divergent term proportional to
$\alpha\ln\gamma$, which is canceled by the corresponding divergence
of the two-body virtual correction in
Eq.~\eqref{eq:two-body-vir}. In the $m_\ell\to0$ limit, however,
$\alpha\to0$, so both this IR-divergent term and the virtual correction
vanish. No virtual correction is therefore required for IR
cancellation in this limit. For finite $m_\ell$, the real and virtual contributions should in
principle be combined to obtain an IR-finite
result~\cite{deLima:2026fsy}, while the finite virtual contribution is
controlled by $\alpha=m_\ell^2/m_P^2$. It becomes relevant
for sizable $m_\ell^2/m_P^2$, such as in
$\pi^+\to\mu^+\bar{\nu}_\alpha\phi$, where the virtual correction to
the corresponding two-body decay $\pi^+\to\mu^+\nu_\mu$ cannot be
neglected~\cite{Dev:2024ygx}.

\bibliographystyle{apsrev4-1}
\bibliography{reference.bib}

\end{document}